\documentclass[unnumsec,webpdf,contemporary,large]{oup-authoring-template}%

\usepackage{amsmath, amssymb}
\usepackage{color, colortbl}
\usepackage{afterpage}
\usepackage{url}
\usepackage{multirow}
\usepackage{rotating}
\usepackage{scalefnt}
\usepackage{hyperref}
\usepackage{ulem}
\usepackage{braket}
\usepackage{longtable}
\usepackage{xr}
\usepackage{soul}
\usepackage{natbib}
\usepackage{gensymb}
\theoremstyle{thmstyleone}%
\theoremstyle{thmstyletwo}%
\theoremstyle{thmstylethree}%

\begin{document}

\journaltitle{PNAS Nexus}
\DOI{DOI HERE}
\copyrightyear{2021}
\pubyear{2021}
\access{Advance Access Publication Date: Day Month Year}
\appnotes{Paper}

\firstpage{1}


\title[]{Structural dynamics of plant-pollinator mutualistic networks}

\author[a]{Aniello Lampo}
\author[b]{Mar\'{i}a J. Palazzi}
\author[b]{Javier Borge-Holthoefer}
\author[b,$\ast$]{Albert Sol\'{e}-Ribalta}

\authormark{Aniello Lampo et al.}

\address[a]{Grupo Interdisciplinar de Sistemas Complejos (GISC),
Departamento de Matem\'aticas, Universidad Carlos III de Madrid, Spain}
\address[b]{Internet Interdisciplinary Institute (IN3), Universitat Oberta de Catalunya, Barcelona, Catalonia, Spain}

\corresp[$\ast$]{ \href{asolerib@uoc.edu }{asolerib@uoc.edu }}

\received{Date}{0}{Year}
\accepted{Date}{0}{Year}


\abstract{The discourse surrounding the structural organization of mutualistic interactions mostly revolves around modularity and nestedness. The former is known to enhance the stability of communities, while the latter is related to their feasibility, albeit compromising the stability. However, it has recently been shown that the joint emergence of these structures poses challenges that can eventually lead to limitations in the dynamic properties of mutualistic communities. We hypothesize that considering compound arrangements --modules with internal nested organization-- can offer valuable insights in this debate. We analyze the temporal structural dynamics of 20 plant-pollinator interaction networks and observe large structural variability throughout the year. Compound structures are particularly prevalent during the peak of the pollination season, often coexisting with nested and modular arrangements in varying degrees. Motivated by these empirical findings, we synthetically investigate the dynamics of the structural patterns across two control parameters --community size and connectance levels-- mimicking the progression of the pollination season. Our analysis reveals contrasting impacts on the stability and feasibility of these mutualistic communities. We characterize the consistent relationship between network structure and stability, which follows a monotonic pattern. But, in terms of feasibility, we observe non-linear relationships. Compound structures exhibit a favorable balance between stability and feasibility, particularly in mid-sized ecological communities, suggesting they may effectively navigate the simultaneous requirements of stability and feasibility. These findings may indicate that the assembly process of mutualistic communities is driven by a delicate balance among multiple properties, rather than the dominance of a single one.}


\maketitle
\section{Introduction}
Mutualistic species interactions --their quantity, diversity, and structure-- play a pivotal role in preserving ecosystems \cite{bascompte2013mutualistic}. However, they are known to respond to varying biotic and abiotic conditions \cite{letten2017linking} and to change across geographic locations \cite{schleuning2012specialization}. Indeed, the impacts of climate change \cite{burkle2013plant,alarcon2008year}, habitat loss \cite{lazaro2022habitat}, and species invasions \cite{traveset2014mutualistic,vizentin2019structure} are profoundly affecting mutualistic interactions all over the planet. As a paradigmatic example of a complex system \cite{newman2018networks}, understanding the influence of these intricate dynamics on system-scale properties presents a major challenge in the field of community ecology. 

In the specific context of plant-pollinator communities, which exhibit marked annual periodicities, it is usually convenient to characterize and analyze these interactions across various time frames \cite{caradonna2021seeing}. This approach is indispensable, as species interactions typically involve complex and intertwined dynamic processes that give rise to diverse patterns at various time scales \cite{cirtwill2018between,tylianakis2018symmetric,burkle2013plant,olesen2008temporal}. At larger time scales (years to decades), turnover and individual interactions vary \cite{olesen2011strong,chacoff2018interaction,cirtwill2018between,alarcon2008year,resasco2021plant} (stochastically, perhaps \cite{macleod2016measuring}), but the overall structure of the interaction networks seems to remain quite stable across seasons \cite{dupont2009spatio,olesen2011strong,alarcon2008year,chacoff2018interaction, resasco2021plant}, also when considering a reasonable range of aggregation windows, from weekly to monthly intervals \cite{schwarz2020temporal}. At shorter time scales, during the course of the season, plant-pollinator communities also undergo changes in the number of species \cite{bramon2020untangling}, leading to variations in connectance \cite{caradonna2020temporal,Suweis2013}, as well as experiencing significant turnover \cite{CaraDonna2017}. These variations in the underlying conditions of the system lead to complex rewirings of the relations between species \cite{CaraDonna2017,caradonna2020temporal,bramon2020untangling} and the emergence of preferential attachment mechanisms \cite{olesen2008temporal} that affect the macro- and mesoscale structure of their interaction network \cite{poisot2015beyond}. 

In each of these temporal scales, scholars have investigated network micro- to macroscopic properties. At the local level, several studies have examined species (node) dynamics and their implications, such as structural roles \cite{bramon2020untangling}, intraday dynamics \cite{bramon2020untangling}, and phenological impacts \cite{Petanidou2014}, among others. 
Moving up to the system's meso- and macro-scale, the emergence of structured interactions has been attributed to various dynamical processes \cite{lewinsohn2006structure,guimaraes2020structure}, including different levels of complexity and detail: niche patterns \cite{saavedra2009simple,santamaria2007linkage}, niche dynamics \cite{Cai2020,nuismer2018coevolution}, eco-evolutionary mechanisms \cite{nuismer2018coevolution,nuismer2013coevolution,thebault2010stability,zhang2011interaction,guimaraes2007interaction,Minoarivelo2016, pinheiro2019new}, phylogenetics \cite{peralta2016merging,chamberlain2014phylogenetic,mello2019insights}, geographical constraints \cite{mello2019insights}, or abundance maximization \cite{Suweis2013}, among others \cite{krishna2008neutral}. While these models have greatly enriched our understanding of ecological communities, elucidating the dynamical properties responsible for the resilience and adaptability of such structured interactions remains challenging, and to some extent limited to time-aggregated interaction networks.

In this last set-up, the seminal work by Bascompte \textit{et al.} \cite{bascompte2003nested} demonstrated that a significant number of plant-pollinator and seed-dispersal networks exhibit nested arrangements \cite{atmar1993measureorder,Mariani2019}, \textit{i.e.} specialist species interact only with subsets of those interacting with the more generalists. Similarly, in \cite{olesen2007modularity} it has been noted that pollination communities can also exhibit a modular character \cite{newman2004finding, newman2006modularity, leicht2008community, fortunato2010community}, with densely linked groups of nodes, which are sparsely connected to the rest of the network. 
From the dynamical point of view, the identification of nested arrangements has been associated with various beneficial properties \cite{pinheiro2019new}, such as promoting diversity \cite{bastolla2009architecture}, maximizing species abundances \cite{Suweis2013}, or enhancing system feasibility \cite{rohr2014structural, Saavedra2015}; which, within the Lotka-Volterra framework, characterizes the range of all possible growth rates that result in positive stationary abundances for all species, given an established interaction matrix.
However, studies suggest that nested networks tend to be less stable \cite{allesina2012stability, staniczenko2013ghost}, whereas greater stabilizing effects are associated with modular structures \cite{stouffer2011compartmentalization, allesina2015modularity, Grilli2016}.

These findings challenge the notion that a single organizational pattern is universally advantageous. Moreover, empirical research \cite{fortuna2010nestedness} and analytical evidence \cite{Palazzi2019Royal} have shown that these two patterns may not be structurally compatible with each other, contradicting the original proposition put forward in \cite{olesen2007modularity}. So far, the study of the interplay between structure and dynamics in ecological communities has primarily focused --with rare exceptions \cite{servan2018coexistence}-- on one-to-one mappings. That is, examining the correspondence of a given dynamical property to a single architectural pattern, or vice-versa. Nevertheless, this view may be too limited to fit some of the complex dynamics observed in natural systems, which may potentially evolve concurrently by optimizing several ecological variables \cite{Tilman1994,McCann2000,ives2007stability,donohue2013dimensionality,kefi2019advancing}, admittedly probably with some correlations \cite{dominguez2019unveiling}. 

A recent direction to fill this gap, currently attracting fresh interest, points at hybrid structural patterns.
These complex organizational configurations, which have also received attention outside ecology \cite{kojaku2017finding}, are defined as a combination of simpler network macro- and mesoscale arrangements at different interacting scales. 
In the field of ecology \cite{lewinsohn2006structure} and bio-geography \cite{presley2010comprehensive}, these have crystalized in the definition of compound structures and, much later, in a formal definition: in-block nested (IBN) networks \cite{sole2018revealing}, that describe communities with compartmentalized species interactions with internal nested organization. 

After the initial definition and identification of these compound patterns in real ecological communities \cite{lewinsohn2006structure,flores2011statistical,flores2013multi}, scholars have focused on exploring their origins as well as in identifying these patterns in other ecological settings \cite{felix2022compound,valverde2020coexistence,diniz2023interplay,queiroz2021bats,pinheiro2022hierarchical}. Several plausible mechanisms for their emergence have been eventually described, {\it e.g.}, niche theory-based \cite{Cai2020,palazzi2021ecological,felix2022framework,felix2022compound,latombe2015beyond}, trait-mediated \cite{Minoarivelo2016}, eco-evolutionary models \cite{pinheiro2019new} (including also simpler ones \cite{valverde2018architecture}), and geographic co-occurrence in combination with phylogenetic constraints \cite{mello2019insights}. 

\begin{figure*}[t] \includegraphics[width=0.95\textwidth]{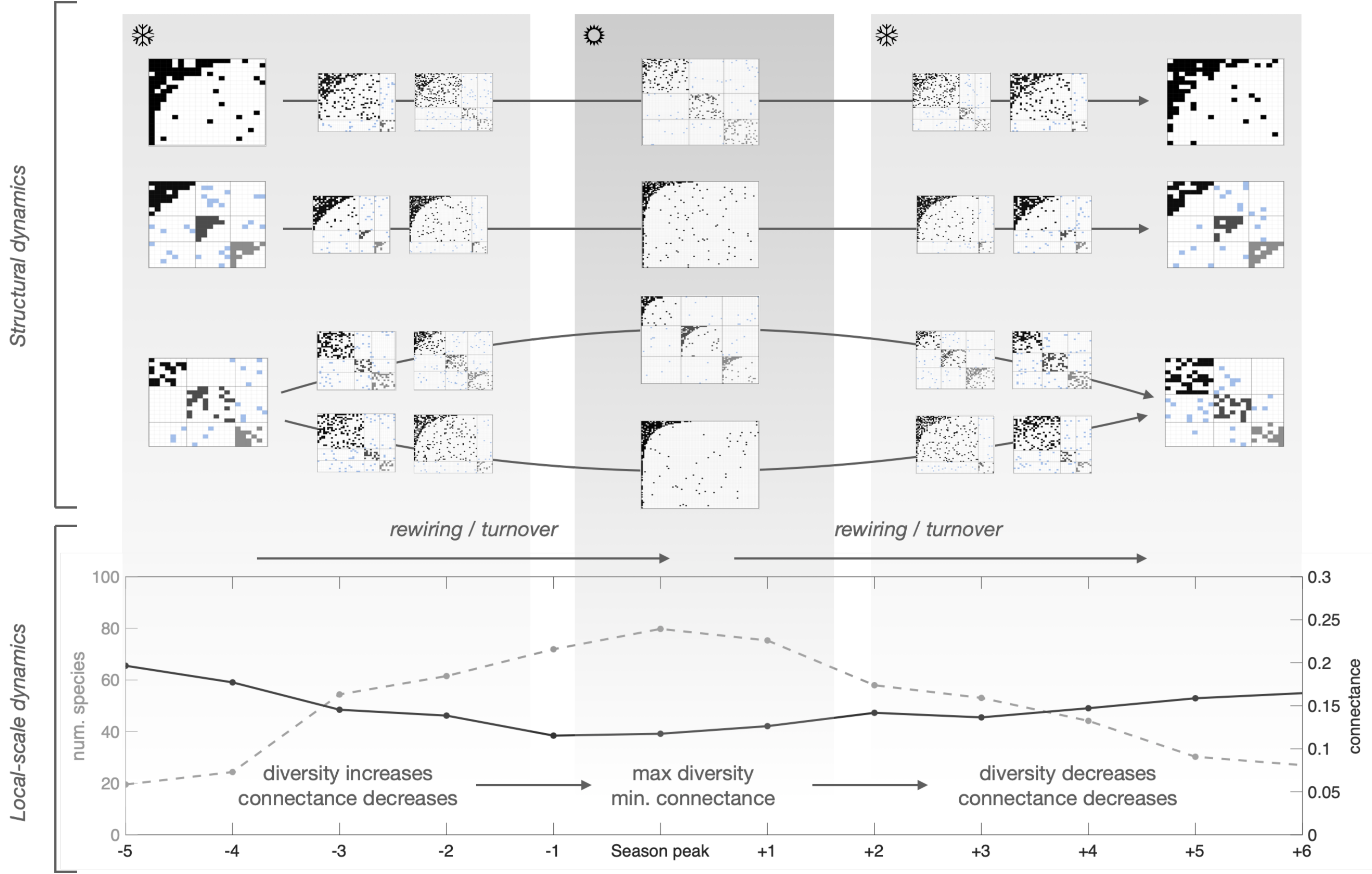}
    \caption{The figure illustrates the hypothesized impact of different pressures and parameter variations (network size and connectance) along the pollinator season on the structural arrangement of plant-pollinator networks. The upper part of the figure illustrates different structural transformations that one might find throughout the pollinator season: nested to modular, compound to nested, etc. The lower part of the figure shows a line plot with the average variation in size and connectance of the interaction networks observed in the empirical dataset used in this paper.}
    \label{fig:intro}
\end{figure*}

In this paper, we attempt to link the temporal analysis of plant-pollinator networks and the existing knowledge derived from time-agnostic aggregated network samples. We probe the structural variability of the interaction network of plant-pollinator communities and confront it to their dynamical characteristics. In agreement with previous literature, we analyze to which extent structural changes of the interaction network are in response to variations of the ecosystem parameters during the pollination season. Additionally, we hypothesize and show that compound structures not only provide ecological communities with dynamic advantages inherited from their constituent building blocks but also that distinct structural arrangements may offer different benefits depending on the system's state. Figure~\ref{fig:intro} illustrates this workflow. Specifically, we analyze the structural characteristics of species interactions throughout the pollination season, identifying a prevalent shift from modular to compound (hybrid) patterns during the peak of the season. These results serve as a motivation for our subsequent theoretical investigation, aiming to confront these arrangements to two dynamical properties: local asymptotic stability (hereafter stability) and feasibility. We explain --in the Lotka-Volterra dynamics framework-- why such transitions may occur on varying size, connectance, and interaction intensity (of both mutualism and competition). The analysis reveals several regimes where nested, modular, and compound structures offer dynamic advantages, while the IBN architecture provides a balancing effect between stability and feasibility for low to mid-sized communities.
We hope that our results offer a new perspective on community assembly, responding to attempts to balance several dynamical properties rather than promoting one over the other.

\section{Results}\label{sec2}

\subsection{Predominant structures in plant-pollinator interaction networks}

To assess the structural dynamics of mutualistic networks, we leverage data from \cite{schwarz2020temporal}, which comprises 30 individual datasets of pollination networks from sites in 9 countries, primarily located in temperate regions. Each dataset tracks the interactions between plant and pollinator species (taxonomic species or morphospecies) within a given time window: daily, weekly, or monthly. Considering that network communities require a minimum density to emerge, we have chosen a monthly time frame for our analysis. Interestingly, despite the potential consequences different aggregation windows may have on network metrics, both the modular structure and nestedness show minimal sensitivity within the range of weeks to months \cite{schwarz2020temporal}.

\begin{figure*}[t] \includegraphics[width=0.95\textwidth]{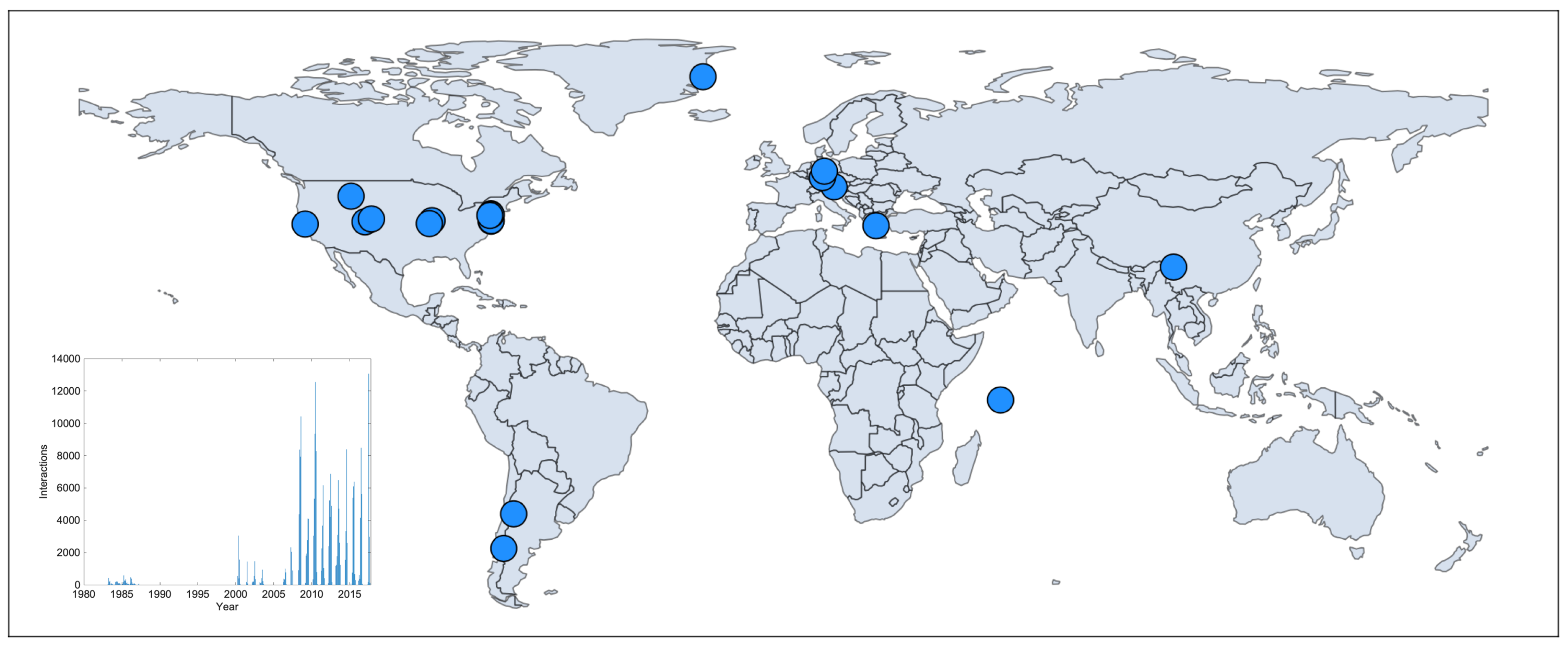}
    \caption{Location of the plant-pollinator communities considered in our study. Inset provides information about the aggregated number of interactions captured over the years in the datasets.}
    \label{fig:map}
\end{figure*}

After several data cleaning procedures (described in Materials and Methods), we selected 20 datasets for analysis, each containing at least three consecutive snapshots corresponding to calendar months. The final dataset includes a community from a tropical region ($-4.7\degree, 55.4\degree$) and one from the Arctic circle ($74.5\degree,-20.6\degree$), while the remaining ones are located in temperate regions. Most tropical and arctic regions were excluded from the analysis due to the limited number of interactions, which hindered a robust temporal analysis. See Figure~\ref{fig:map} to visualize the locations of the plant-pollinator communities considered in our study.

For each of these datasets, we quantified the degree of nestedness, group structure, and in-block nestedness using specific measures. The NODF-like measure, as defined in \cite{sole2018revealing}, was used for nestedness evaluation, while modularity was optimized using the extremal optimization algorithm \cite{duch2005community}. In the case of in-block nestedness, we employed the optimization method defined in \cite{sole2018revealing}. For specific details about these measures, we refer the reader to Sec.~S1 in the Supplementary Materials.

It is important to note that the obtained values for the different structural organizations cannot be directly compared due to methodological differences \cite{guimera2004modularity,simmons2019beware}. We opted to compare them by assessing their statistical significance by means of z-scores, which were obtained by performing 150 randomizations of each network. These randomizations were generated using a corrected version of the null model proposed in \cite{bascompte2003nested}, which respects the network degree of the plant-pollinator bipartite networks. In other words, the marginals of rows and columns of the network's adjacency matrix are preserved. For detailed information about the null model, the randomization process, and their performance, please refer to Section S2 of the Supplementary Materials.

\begin{figure*}[t]
	\begin{tabular}{l}
            \includegraphics[width=0.95\textwidth]{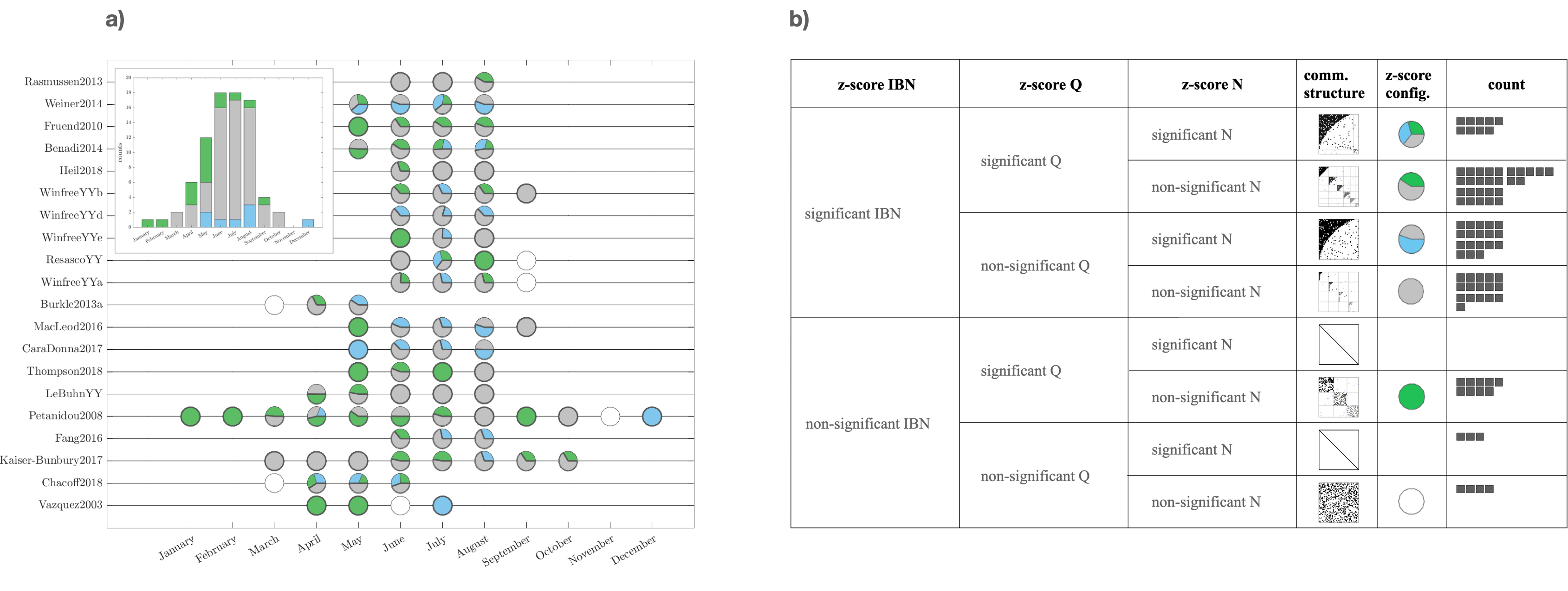}
    \end{tabular}
    \caption{\textbf{(a)}. Temporal evolution of the interaction networks' structure in 20 plant-pollinator communities. Green, gray, and blue colors in the pie diagrams relate to the significant z-scores ($\mbox{z-score} > 1.96$) obtained for modular, hybrid (in-block nested), and nested arrangements, respectively. Slices displaying a single color indicate that only one structure was found to be significant. In slices containing multiple colors, the area of each color represents the respective proportions of the different z-scores obtained. The area outlined with a thicker line indicates the structural descriptor that achieved the highest z-score. White circles indicate that no structure was found to be significant. Datasets are sorted in descending order by latitude, from North to South. The inset aggregates the results by month, highlighting the most predominant structure (indicated by the highest z-score) across the various temporal slices of the dataset. \textbf{(b)} provides an intuitive hierarchy, matching each combination of significant z-scores to a prototypical structural pattern. Column 4 shows the expected structural configuration, while Column 5 relates the analysis to panel (a) of this figure. Lastly, Column 6 indicates the frequency of each structure detected in the dataset. Section S3 in the Supplementary Material provides the raw values from the structural analysis. Also, it includes a scatter plot that illustrates the relationship between the z-scores obtained for $Q$ and $\mathcal{I}$ arrangements.}
    \label{fig:real_nets_structural}
\end{figure*}

The results are provided in Fig.~\ref{fig:real_nets_structural}. Panel (a) shows that most of the communities achieve a significant z-score for more than one structural organization (multicolor pie diagrams) and these vary throughout the pollination season. This supports our hypothesis that structural transitions are common during these periods. Panel (b) is a support guide for our results, mapping each configuration of significant z-scores to prototypical structural arrangements. It is interesting to examine the six possible configurations and their frequencies in our dataset, as well as the two configurations that are not possible because nested networks are inherently IBN networks with a single compartment. Although implausible, we identify three situations where the nested arrangement is the only significant outcome, which we attribute to a suboptimal solution of the IBN-maximization process.

In the results, few communities display all three structural types simultaneously, as nested arrangements tend to impede the emergence of pure modular structures (row 6 in Fig.~\ref{fig:real_nets_structural}b) \cite{Palazzi2019}. These configurations are only feasible in heterogeneous-sized nested modules, provided that at least one module is large enough to facilitate system-scale nestedness and several others are sufficiently dense to maintain significant modularity. It is also interesting to analyze the relatively common occurrence of IBN structures that are neither modular nor nested (gray-only circles in Fig.~\ref{fig:real_nets_structural}). These arise when modules are structured in a nested form but lack sufficient density to achieve significant modularity, usually because the nested structure is very stylized. Nested and modular patterns coexisting with IBN structures at different degrees are quite common since this compound arrangement incorporates elements of both. Significant IBN communities that also achieve significant $z_Q$, due to their modular organization, are the most common pattern. Finally, pure modular structures (green circles in Fig.~\ref{fig:real_nets_structural}) are rare and typically emerge at the beginning of the sampling periods in the datasets.

Examining the temporal evolution within each dataset, we find that most of them undergo transitions between structural patterns (z-score configurations vary). Although it's crucial to contextualize the results to the particularities of each mutualistic community, our analysis yields several conclusions.
The IBN arrangement, specifically featuring more than one community and consistently obtaining higher z-scores, is the most prevalent configuration during the peak of the pollination season. This is corroborated by the inset panel of the figure and the hierarchical classification shown in panel (b). Moreover, the frequent co-existence of IBN with other patterns in multiple datasets suggests complex dynamics, involving variation in both size and internal organization. For example, several datasets incur in transitions from modular to IBN structures (\textit{e.g.}, Fruend2010, Thompson2018, Petanidou2008; or Heil2018 and LeBuhnYY to a lower degree), and vice versa (\textit{e.g.}, Kaiser-Bunbury2017, Rasmussen2013), indicating a reorganization within their community structures. Transitions between different degrees of IBN and nestedness are also common (\textit{e.g.}, WinfreeYYd, MacLeod2016, and CaraDonna2017), suggesting interesting dynamics of growth-split and shrinkage-merge of a large nested block.

\begin{figure}[t]
	\begin{tabular}{l}
            \includegraphics[width=0.45\textwidth]{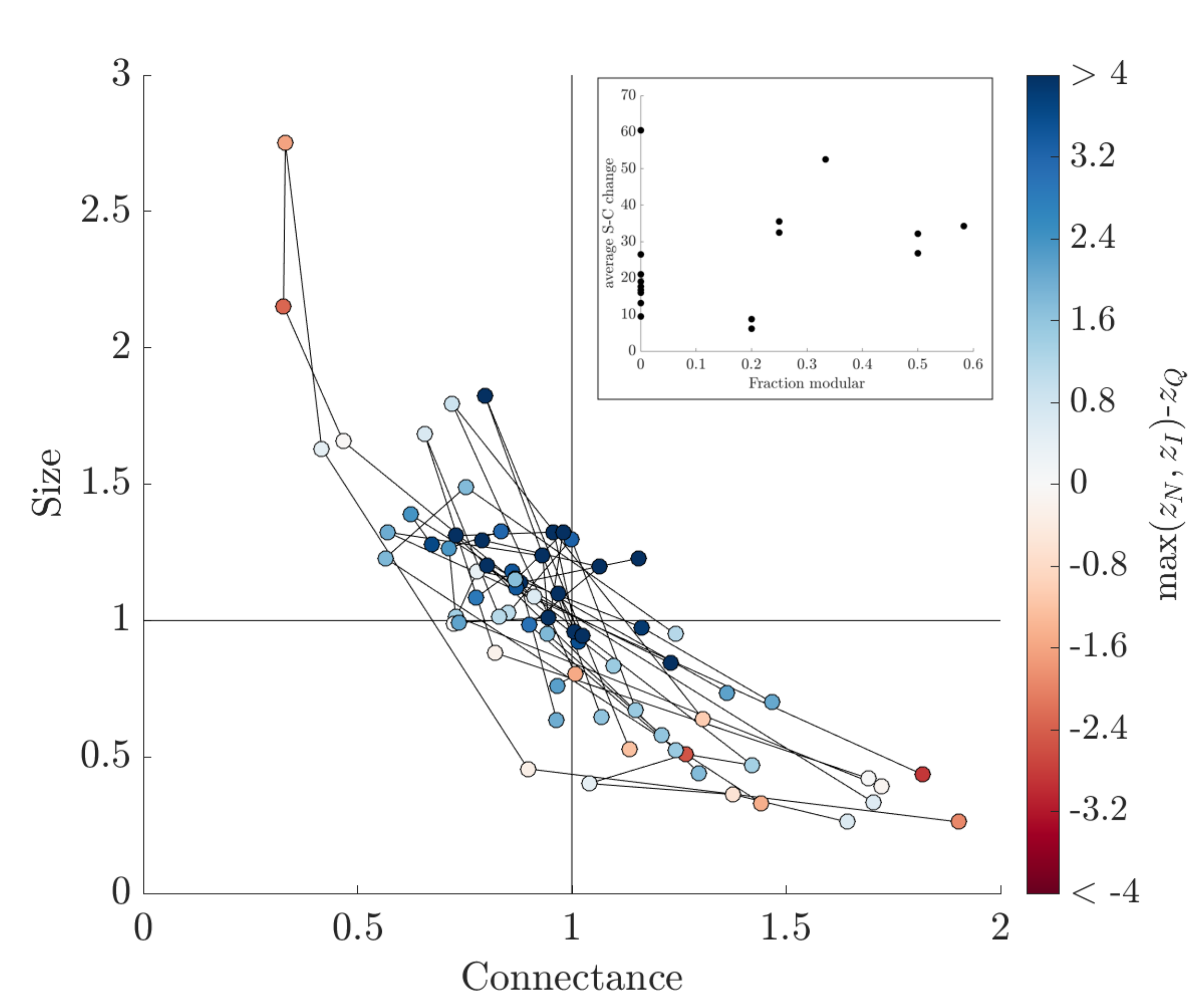}
    \end{tabular}
    \caption{Diagram illustrating the evolution of size and connectance in relation to the differences in statistical significance of the analyzed structural arrangements. Each point represents a snapshot of the interaction network from a specific dataset, with consecutive snapshots connected by lines. The color scale indicates the absolute difference between the z-scores obtained for nested-like and modular structures. The inset displays the average displacement of consecutive temporal snapshots over the size-connectance diagram with respect to the fraction of time a modular structure is found to be predominant in the plant-pollinator community. Each point corresponds to a dataset. See Sec.~S3 of the Supplementary Material for the raw values obtained in the analysis.
    }
    \label{fig:sc_diagram_structuralTransitions}
\end{figure}

To complete the analysis, we assess the impact of variations in size and connectance on the observed structural transitions, we explore the z-scores for nestedness, IBN, and modular structures across the size-connectance diagram. The results, displayed in Fig.~\ref{fig:sc_diagram_structuralTransitions}, show that communities with sizes and connectance values near their averages commonly exhibit one or more (IBN) nested modules. However, when communities deviate from their average size and connectance values, they exhibit predominantly pure modular structures with fewer nested patterns. The inset shows a scatter plot depicting the relationship between the fraction of temporal networks predominantly exhibiting a modular structure and the average displacement (represented by the average Euclidean distance between consecutive temporal snapshots) across the size-connectance diagram (main plot). While there appears to be a positive trend, it is important to note that the significance of this relationship cannot be determined due to the limited amount of available data.

\subsection{Synthetic experiments}

\begin{figure*}[t!]
  \begin{center}
  \includegraphics[width=0.99\textwidth]{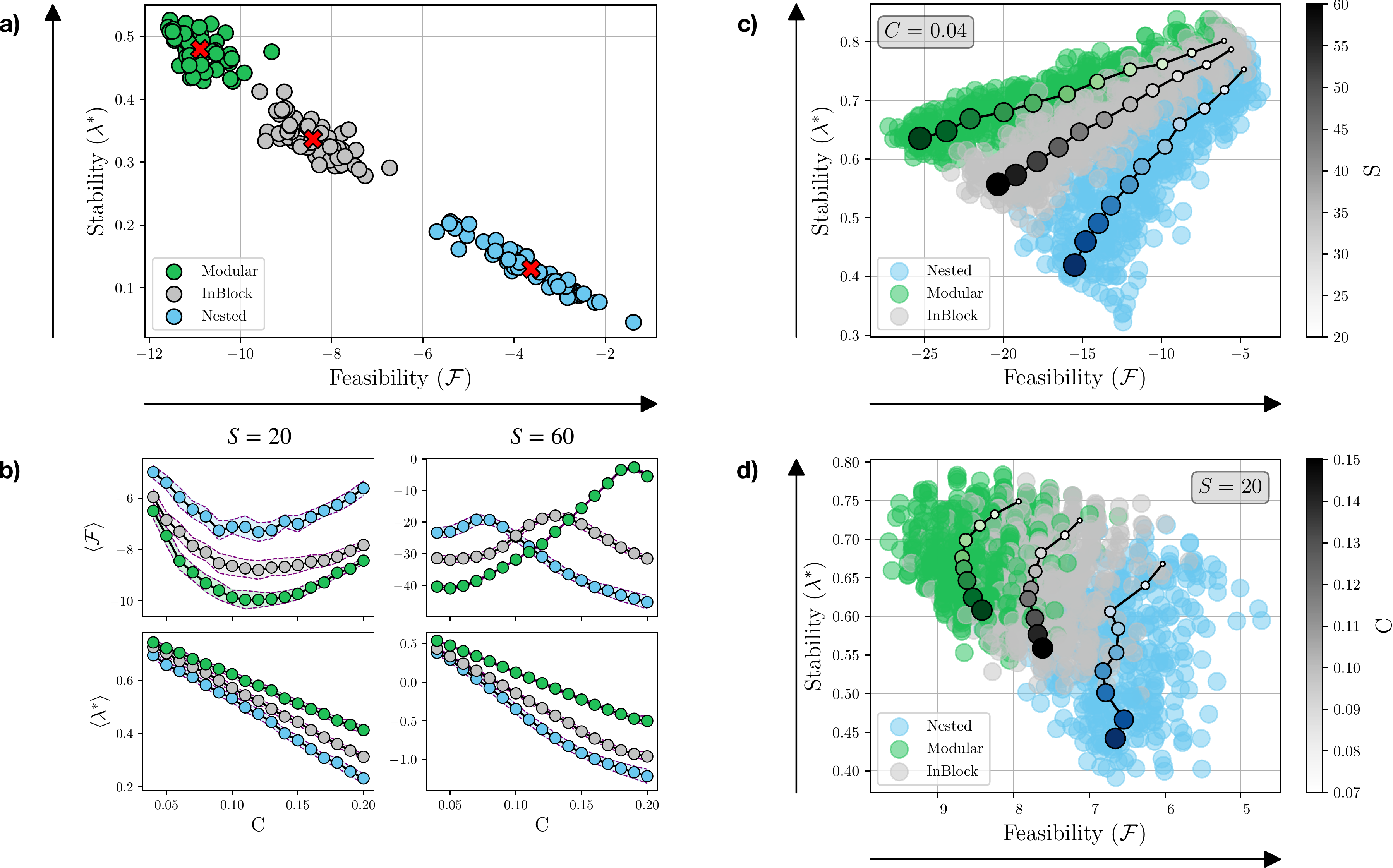}
  \end{center}
  \caption{\label{fig:FeasVsStab} Stability and feasibility performance in nested, modular, and in-block structured networks. 
  \textbf{(a)} Stability vs. feasibility analysis for an ensemble of synthetic networks with varying levels of in-block nested ($B=2$, $p\in [0,0.06]$, $\mu\in [0,0.06]$,  $\xi\in [1.2,1.5]$), nested ($B=1$, $p\in [0,0.1]$, $\mu=0$, $\xi\in [1.85,2.55]$), and modular ($B=2$, $p=1$, $\mu\in [0,0.1]$,  $\xi\in [1.1,1.5]$) features, with $S=20$ species ($|A|=|P|=S$) and connectance $C=0.2$. Each point represents a network, with its color indicating the type of structural arrangement it contains: blue for nested networks, green for modular networks, and gray for in-block nested networks. Red crosses are located at the average values of stability and feasibility of the corresponding network clusters. \textbf{(b)} Stability and feasibility dependence on connectance (C) for nested, in-block and modular networks of size $S=20$ (left) and $S=60$ (right). For a fixed connectance, the average feasibility and stability on several network realizations are reported (shadowed areas represent the variance). All the experiments are performed with $\gamma = 0.1$ and $\omega = 0.01$ parameters of the Lotka-Volterra dynamics. \textbf{(c)} Here, the stability-feasibility ordering across architectural patterns is portrayed for different values of connectance, varying in the range $\left[0.07,0.15\right]$ at fixed size. Dot-solid lines depict the average of the related network ensemble with size, while darker colors indicate increasing connectance. Deviation around central values arises because of the noise introduced in the interactions between species (see Materials and Methods). \textbf{(d)} Along the same line, the information of the previous plot is presented for different values of the size $\left[20,60\right]$  at fixed connectance.}
\end{figure*}

The analysis of mutualistic networks during the pollination season has revealed that the structure of interaction networks varies, potentially due to changes in size and connectance over time. Due to the dissimilar characteristics of the datasets, \textit{e.g.}, variations in mutualistic and/or competition strengths, climate regions, or sampling effort, a direct comparison between them is not advisable. Therefore, we resort to synthetic experiments to further investigate the relationship between the network's structural arrangements and their dynamical response. 
Specifically, we examine the performance of the three prototypical structural arrangements (pure modular, nested, and IBN) on synthetically engineered mutualistic communities and we evaluate their impact on two commonly studied dynamical characteristics: stability and feasibility, within the Generalized Lotka-Voltera dynamics framework. See Materials and Methods for all the definitions and details on the generative network model. 
Results are presented in the different panels of Fig.~\ref{fig:FeasVsStab}. Panel (a) shows the results obtained for a particular size and connectance configuration. It is apparent to the naked eye that modular networks (green dots) are the most stable ones, while nestedness (blue dots) penalizes stability. In this sense, stability defines an ordering in which modularity is on the top, nestedness at the bottom, and in-block nested structures (grey dots) lay between those. On the x-axis, feasibility shows the reversed behavior: nested architectures show the highest values of $\mathcal{F}$, while modular ones fall behind. Importantly, in both situations, IBN offers intermediate dynamical properties. More formally, we have:
\begin{equation}\label{eq:hierarchies}
\langle\mathcal{F}_{\mathcal{N}}\rangle>\langle\mathcal{F}_{\mathcal{I}}\rangle>\langle\mathcal{F}_{Q}\rangle,\quad\quad \langle\lambda^*_{\mathcal{N}}\rangle<\langle\lambda^*_{\mathcal{I}}\rangle<\langle\lambda^*_{Q}\rangle,
\end{equation}
where $\langle\mathcal{F}_{\mathcal{X}}\rangle$ ($\langle\lambda^*_{\mathcal{X}}\rangle$) indicates the average value of feasibility (stability) over the ensemble of synthetic networks with predominant $\mathcal{X}$ structure.

Panels (c) and (d) show the same information as the previous plot, varying the size and the connectance, respectively. These suggest that the observed relationship in panel (a) is, to a large extent, robust to variations in the size and connectance of the mutualistic community. In this sense, our results regarding pure nested and pure modular communities agree with the previous literature, where nestedness was found to hinder stability \cite{allesina2012stability,staniczenko2013ghost} favoring feasibility \cite{Saavedra2015}; and modularity was found to boost stability \cite{allesina2015modularity}, while its effects on feasibility remained, to the best of our knowledge, untested so far. From the mathematical point of view, the ordering provided by stability may be understood in terms of the Gershgorin theorem \cite{Varga2004}, linking the real part of the largest eigenvalue with the row sum of the interaction matrix, which is maximized in nested networks because of the presence of generalist species (see Sec.~S4 of the Supplementary Material). This is certainly true for pure mutualistic systems \cite{allesina2012stability} but, worth highlighting, these results seem to be robust for an average competition strength up to $10\%$ of the mutualistic one. 

The case of feasibility suggests varying levels of complexity depending on the network parameters. In small communities (S=20), panel (b) demonstrates consistent behavior across all connectance levels. However, for larger networks (S=60) complexity increases, marked by the non-monotonic behavior of $\langle\mathcal{F}\rangle$ as connectance values rise. This phenomenon may find its roots in the relative difference between effective and critical competition, as it was shown for structural stability in \cite{pascual2017mutualism}. Nevertheless, this non-monotonic dependence requires further analysis, which is beyond the scope of this work. Whichever the underlying cause of this behavior, it is easy to see that the three structures become optimal $\langle\mathcal{F}\rangle\mbox{-wise}$ within some parameter range: nested structures are optimal for $C\in{(0.04,0.10]}$, IBN structures for $C\in{(0.10,0.15]}$ and modular structures for $C\in{(0.15,0.20]}$ when $S=60$. These results unveil the existence of three regimes, in which (a) IBN offers a trade-off between feasibility and stability, (b) IBN offers advantages concerning feasibility, and balance on stability and (c) modular networks maximize both feasibility and stability, being in that situation indisputably the best interaction pattern. In the other two of these regimes, IBN structural arrangements are more beneficial than any of its non-hybrid counterparts, balancing their dynamical properties as we hypothesized. To wrap up these results, see also Figs.~S5, S6, and S7 in Sec.~S5 of the Supplementary Material, where we report results for pure mutualistic scenarios and varying the number of species $S$.

Until now, our analysis has been conducted with moderate mutualistic strength, $\gamma=0.1$, and a low competition regime, $\omega = 0.01$, which is approximately an order of magnitude lower than the mutualistic strength. Both parameters are in agreement in order of magnitude with the values used in the literature regarding the analysis of stability and feasibility \cite{bastolla2009architecture, allesina2012stability, Suweis2013, Saavedra2015, pascual2017mutualism, Gibbs2018}.

\begin{figure}[t] \includegraphics[width=0.5\textwidth]{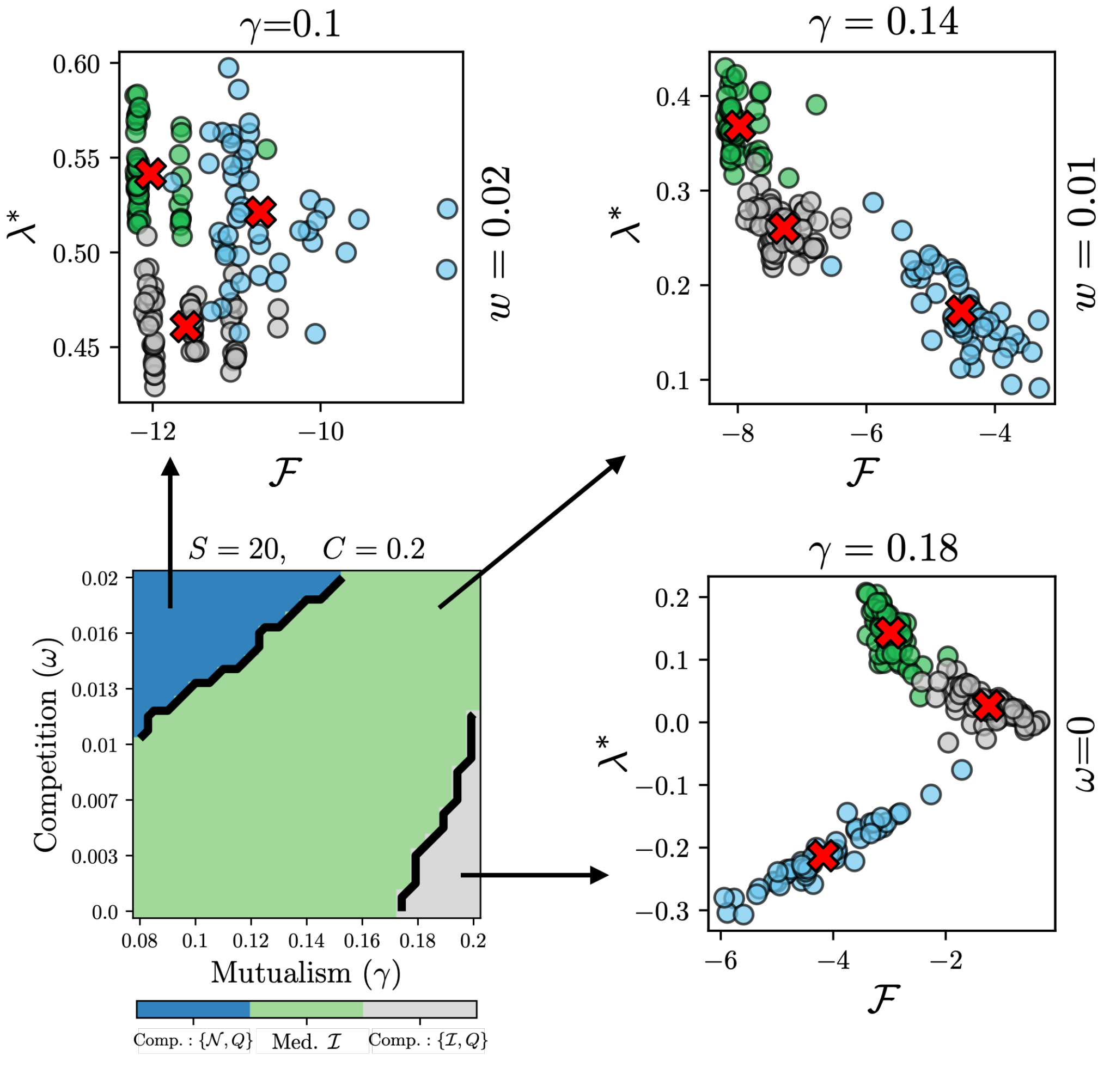}
    \caption{Validity of the mediating role of IBN as a function of mutualism and competition strength. The green area depicts the parameter regime where Eq.~\ref{eq:hierarchies} holds, {\it i.e.}, the mediating role of IBN structures is present. Instead, the blue (grey) area indicates the parameter range where nested (modular) structures are best at balancing stability and feasibility. Experiments were conducted with networks of $S = 20$ and $C = 0.2$.}
    \label{fig:comp_mutualism_diagram}
\end{figure}

To deepen our understanding of this aspect, we explore the relationship between the mutualism and competition parameters ($\gamma$ and $\omega$) of the Generalized Lotka-Volterra model and the dynamical properties linked to the different structural arrangements. Figure~\ref{fig:comp_mutualism_diagram} displays the different situations we find over the $\gamma-\omega$ space. We identify three regions depending on the stability and feasibility ordering. The green region is such that the mediating role of IBN, as expressed in Eq.~\ref{eq:hierarchies}, is preserved. Such ordering prevails across weak and moderate $\gamma$ and $\omega$ values, but changes occur outside that wide range. Specifically, in the limit of large competition, IBN arrangements become the least stable (blue region in Fig.~\ref{fig:comp_mutualism_diagram}). In this scenario, nested arrangements (which can also be viewed as IBN networks with a single block) emerge as the most feasible, competing closely with modular arrangements in terms of stability. In contrast, in cases where mutualism greatly outweighs competition strength (the gray region in Fig.~\ref{fig:comp_mutualism_diagram}), there is an impact on the observed feasibility ordering (as defined in Eq. \ref{eq:hierarchies}). In these scenarios, networks exhibiting IBN structures tend to be the most feasible, displaying also stability values similar to those of modular structures. Figure~S8 in Sec.~S6 of the Supplementary Material provides further details on the transition between these three regions.

\section{Discussion}\label{secdiscussion}

\subsection{Seasonal structural dynamics of plant-pollinator networks}
Seasonal dynamics of pollination networks at the intra-season scale are frequently investigated in terms of three phases: assembly, intermediate dynamics, and disassembly \cite{bramon2020untangling}. While various species interaction-level mechanisms have been identified (\textit{e.g.}, turnover, interaction rewiring), their impact on the meso- and macroscopic structures remains largely unexplored. These mechanisms served as the foundation for designing iterative models that can capture properties at the system scale. Preferential attachment (and detachment) mechanisms have been found to mimic community assembly and disassembly at seasonal scale. As new species enter the community, they tend to connect to the most generalist \cite{olesen2008temporal}, probably biased by some fidelity mechanism \cite{macleod2016measuring}. Similar models in other fields \cite{Mariani2019,konig2014nestedness}, \textit{e.g.}, those attempting to maximize the individual centrality, have indeed shown to promote the emergence of nested structures. 

In our analysis of species interaction data, we observed that pure modular structures predominate over nested configurations during the early stages of assembly periods, from January to May. Although this seems contradictory to the results in Fig. \ref{fig:FeasVsStab}, which show that modular structures are less feasible compared to all other arrangements in small communities, our theoretical analysis suggests that this scenario is plausible in environments with low competition and large mutualistic parameters (as depicted in the grey area of Fig.~\ref{fig:comp_mutualism_diagram}). In such conditions, modular arrangements can achieve greater feasibility and stability than nested configurations and can closely compete with IBN structures (refer to Fig.~S8 in the Supplementary Material).

The detected non-monotonic implications of the competition and mutualism parameters on the Generalized Lotka-Volterra and their consequences regarding stability and feasibility properties deserve thorough investigation to gain understanding of these complex behaviors. In this work, we have explored a comprehensive range of parameters. However, there is generally a limited direct connection between theoretical models and their practical uses, especially when measuring competition in real mutualistic systems. To better understand mutualistic communities, we need to narrow the gap between theoretical ecology and fieldwork. Additionally, the non-linear dependence of feasibility on network size and connectance may also play a crucial role in the assembly of mutualistic communities, especially considering the crossover in the feasibility diagram (see Fig.~\ref{fig:FeasVsStab}). This crossover point could signify a critical transition in the community dynamics affecting the overall stability and persistence of the ecosystem. A deeper exploration of these intricate relationships is essential for a holistic understanding of the underlying mechanisms shaping mutualistic interactions in ecosystems.

Intermediate dynamics, occurring around the peak of the season, are often overlooked in specific studies (with exceptions in \cite{bramon2020untangling}), but interactions in such stages may bear resemblance to those observed in aggregated interaction networks. Microscopic models \cite{Cai2020,palazzi2021ecological} suggest that hybrid structures may emerge through an abundance-maximization process \cite{Suweis2013} atop niche-structured population dynamics. However, the link between these compound structures and their global dynamical properties remains largely unexplored, with existing knowledge limited to their constituent building blocks \cite{staniczenko2013ghost,Thebault2010,rohr2014structural,Saavedra2015,liu2023feasibility}. Our work contributes to expanding the field in this aspect, shedding light on the intricate interplay between different structural arrangements in mutualistic communities. Lastly, we emphasize the significance of comprehending the properties maintained by mutualistic communities beyond their pairwise interaction mechanisms. While the study of link dynamics provides valuable insights into the functioning of mutualistic communities, understanding the system-scale dynamical properties is crucial for gaining a comprehensive understanding of mutualistic communities and ensuring their long-term viability in the face of changing ecological conditions.

\subsection{The mediating role of hybrid structures.}

Despite the progress made by network theory in efficiently detecting and measuring network patterns, community ecologists are still studying the relationship between these observed arrangements and the dynamic properties of ecosystems \cite{Okuyama2008,Thebault2010,allesina2012stability,stouffer2011compartmentalization,Suweis2013,staniczenko2013ghost,rohr2014structural,Saavedra2015,allesina2015modularity,Grilli2016,pascual2017mutualism}. 

To offer a new perspective to the stability-feasibility debate \cite{Saavedra2015}, in this paper we looked into compound structures with the hypothesis that they may inherit beneficial dynamical properties from their building blocks which, in turn, may help ecological communities to persist in time. We consider the in-block nested configuration, where nestedness and modularity interfere at a network’s mesoscale, and show that may provide a mediating role between stability and feasibility. 

Our results have important consequences on the mechanisms governing the organization of mutualistic communities. So far, the question has been addressed as finding the key property shaping ecosystems assemblage \cite{Capitan2009,Campbell2011,Saavedra2015,Servan2021,carpentier2021reinterpreting}. Herein, analyzing communities only in terms of nestedness could lead for instance to the conclusion that these promote feasibility over stability, but not both at the same time. A similar reasoning applies to the pair stability-modularity. The introduction of an intertwined architecture permits us to revisit and deepen this finding. Particularly, the emergence of in-block nestedness in real communities, associated now with a trade-off between stability and feasibility, paves the way to the hypothesis that the fundamental criterium underlying the assembly process is the equilibrium between those (and possibly more) properties, rather than the predominance of one of them. In this work, we validate this statement by means of examining two specific fundamental structural patterns (and their derived compound one). Nevertheless, the literature has explored several others, such as gradient \cite{lewinsohn2006structure} or core-periphery \cite{martin2020core, miele2020core}, among others.

The study over a large ensemble of synthetic networks illustrates that the mediating role of IBN is beneficial for ecological communities over a wide range of parameters, but depends on the connectance and size of the community. In small and very sparse communities, nestedness may suffice to guarantee a feasible and, to some extent, stable system. As communities get larger and denser, we detect an unexpected effect consisting of a reversing of the feasibility ordering. Only in this regime, both stability and feasibility are promoted by modularity, which stands out as the optimal pattern for ecosystem assemblage. While it may be relatively uncommon to find empirical networks within the modular-optimal range, due to the negative correlation between community size and connectance, our simulations indicate that these transitions occur in regimes of size and connectance that align with empirical network characteristics. All in all, our results indicate that there may be different structural adaptations that can serve the need of mutualistic communities to be both feasible and stable, as they evolve into larger (smaller) or denser (sparser) systems. This idea can enhance our insight into ecosystem assembly processes, where an optimal size-connectance-architecture relationship may be relevant.

\section{Materials and methods}\label{sec:methods}

\subsection{Generalized Lotka-Volterra dynamics}
The study of ecological communities is typically based on the analysis of species interaction networks, where nodes represent species and edges reflect the type and strength of interactions between them. In the adjacency matrix $M_{ij}$, a link is turned on if species $i$ and $j$ interact.

Based on this interaction network, species abundances can be mapped to a set of time-dependent functions $x_{i}(t)$, and their temporal evolution is commonly studied with the Generalized Lotka-Volterra model:
\begin{equation}\label{GeneralizedLotkaVolterraEq}
\dot{x}_{i}=x_{i}\left(r_i-\sum_{j}{ M}_{ij}x_j\right), 
\end{equation}
where the indexes $i$ and $j$ run over the system species and parameters $r_i$ indicate the intrinsic growth rate coefficients, ruling the dynamics of the $i$-species when interspecific interaction is dropped out. 

In bipartite mutualistic communities ({\it e.g.}, plant-pollinators or seed-dispersal), we categorize species into two distinct groups, $A$ and $P$. For the sake of simplicity, during our analyses, we will assume that both sets of species, $A$ and $P$, have the same size $|A| = |P| = S$. Relations between species in different groups are assumed to be mutualistic, and competitive within groups. For these bipartite networks, the adjacency matrix describing species relation exhibits a particular shape, 
\begin{equation}
M = \begin{pmatrix}
\Omega_{AA} & -\Gamma_{AP} \\
-\Gamma_{PA} & \Omega_{PP} 
\end{pmatrix}.
\end{equation}
Block $\Omega^{AA}$ $\left(\Omega^{PP}\right)$ represents the competitive interactions between species corresponding to the set $A$ ($P$), and the block $\Gamma^{AP}$ describes the mutualistic interaction between species corresponding to different sets. Both interaction matrices, $\Omega$ and $\Gamma$ may have a particular structure or else be unstructured (random). 

\subsection{Temporal Segmentation of Plant-Pollinator Interaction Networks}

Plant-pollinator interaction data compiled in \cite{schwarz2020temporal} offers a solid background to study the structural change of interactions in mutualistic communities, but it requires some cleaning effort. Each dataset records the raw number of interactions per species pair, without making any additional assumptions (\textit{e.g.}, estimating unobserved interactions), and these records are gathered at different sampling frequencies: the most common is on a daily basis (963 days in the period 2000-2017), followed by a significant number of weekly records (161 weeks in the period 1888-2015), and monthly records (62 months in the period 1983-2013). In total, the datasets include $256\cdot 10^3$ interactions. The inset of Fig.~\ref{fig:map} shows the monthly distribution of these interactions over time.

\begin{figure}[t]
	\begin{tabular}{l}
            \includegraphics[width=0.45\textwidth]{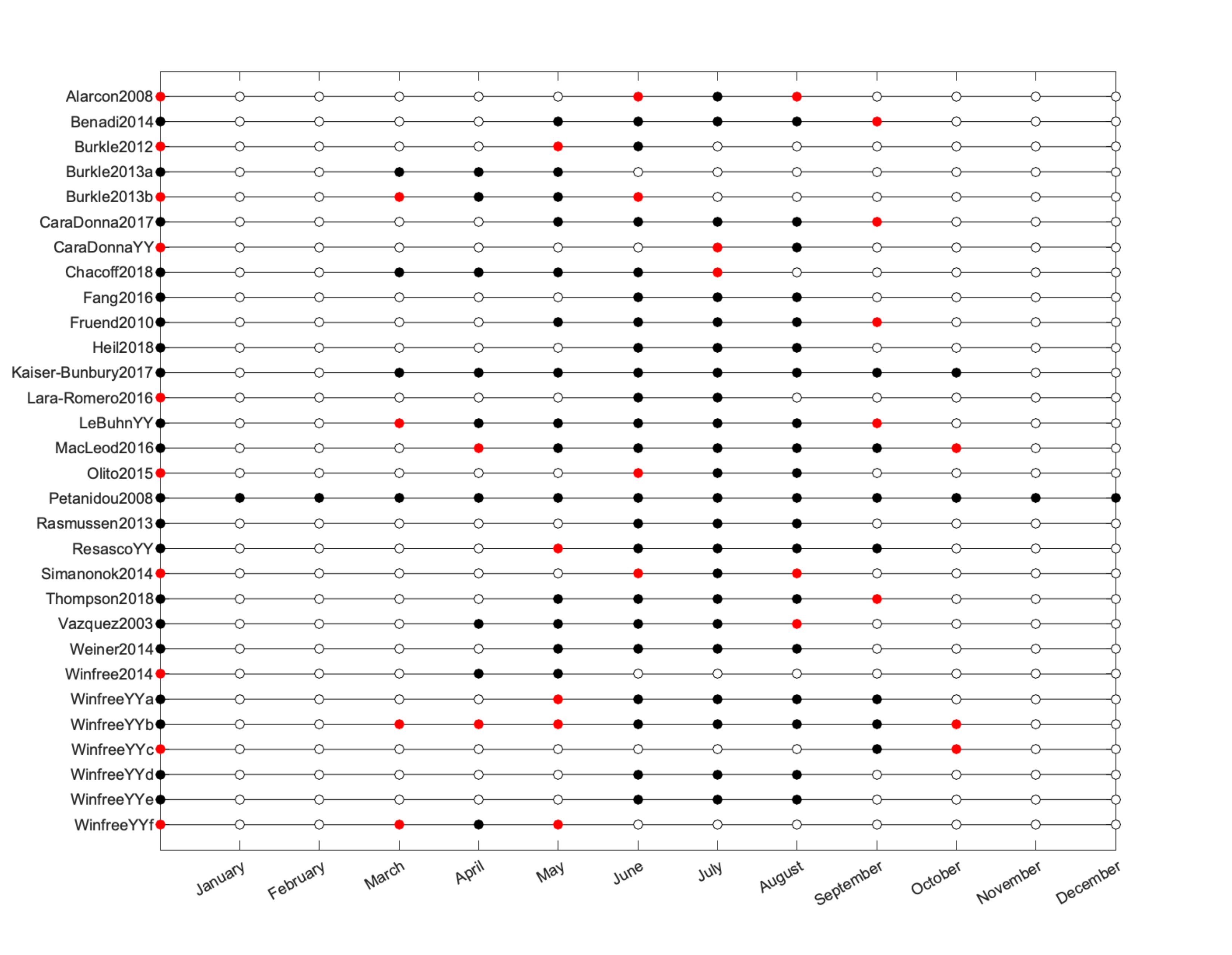}
    \end{tabular}
    \caption{Description of the temporal slices used to construct the interaction network for each dataset in \cite{schwarz2020temporal}. Red dots indicate slices and datasets (dots near the name) we discard because lack of data to assemble the interaction network.}
    \label{fig:discard_slice}
\end{figure}

To integrate all datasets into a common temporal dimension, we opted to collect interactions by calendar month. However, due to potential misalignment with field experiment dates, we conducted a visual inspection and discarded slices where the number of days between the first and last sampling was much lower than the number of days in the respective month. For example, the June slice of Alarcon2008 was discarded since the first sampling was on 17-Jun-2003 and the last one was on 28-Jun-2003. We expect these decisions to help minimize the bias caused by differences in sampling effort. Furthermore, since pollinators could be determined by taxonomic species or morphospecies, we chose to keep the most specific name. After implementing all these procedures, we discarded any dataset that did not have at least 3 consecutive correct temporal slices. Figure~\ref{fig:discard_slice} shows the discarded temporal slices and datasets.

\subsection{Synthetic network generation}

\begin{figure}[t!]
    	\begin{tabular}{lll}
                {\bf (a)} & {\bf (b)} & {\bf (c)}
                \\
                \includegraphics[width=0.3\columnwidth]{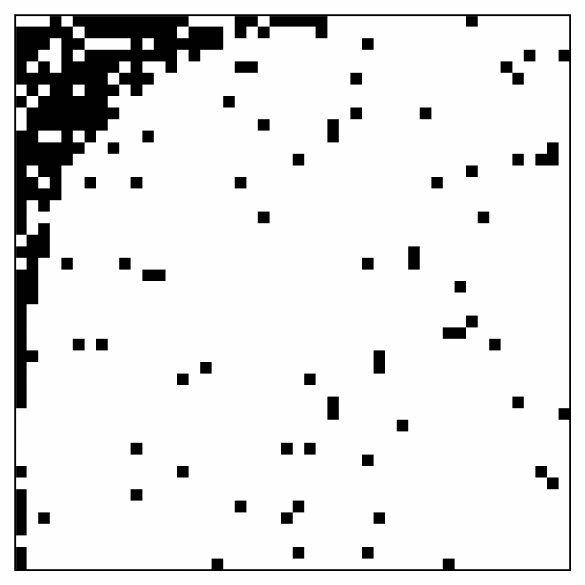}
                &
                \includegraphics[width=0.3\columnwidth]{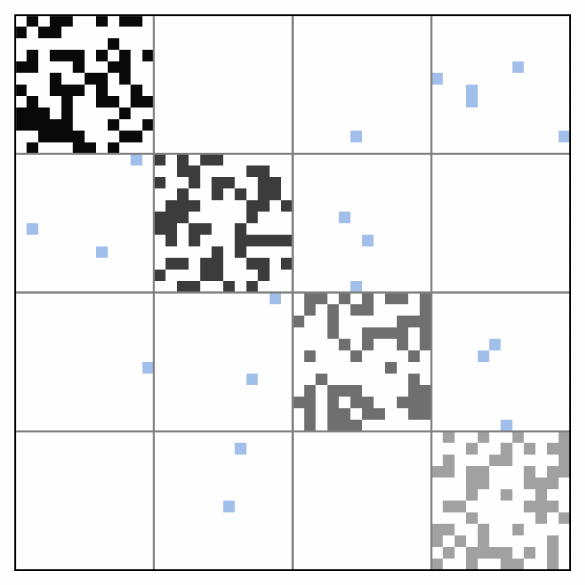}
                &
                \includegraphics[width=0.3\columnwidth]{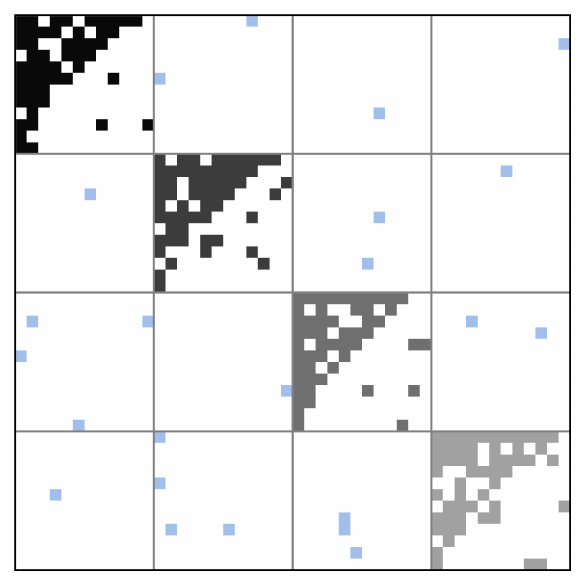}                
        \end{tabular}
  \caption{
Examples of adjacency matrices of networks associated with nested (a) modular (b) and IBN (c) structures. Rows represent species of group $A$ and columns represent species of group $P$. Matrix entries portray the mutualistic interaction links between the two groups. In nested networks, the specialist species interact only with sub-sets of species interacting with the more generalists, and the related adjacency matrices manifest the traditional triangular structure. Modular networks are composed of weakly interlinked groups of species (modules) with strong internal connectivity, and this yields an adjacency matrix divided into blocks. IBN matrices are divided into blocks with an internal nested structure. 
\label{fig:IBNMat}} 
\end{figure} 

Aligned with the objectives of our paper, we focus on examining the dynamical responses of three distinct structural arrangements: modular, nested, and in-block nested. These arrangements are reflected in the off-diagonal mutualistic blocks of the species interaction adjacency matrix $M$, denoted as $\Gamma^{AP}$ and $\Gamma^{PA}$. Meanwhile, the diagonal blocks of matrix $M$ are considered unstructured and assumed to be fully connected, in line with state-of-the-art studies \cite{allesina2012stability,Suweis2013,Saavedra2015}. 

To construct a diverse synthetic ensemble that encompasses the various structural patterns of $\Gamma^{AP}$ under consideration, we find it advantageous to employ the model introduced in \cite{sole2018revealing,Palazzi2019}. This model naturally spans nested, modular, and in-block nested (IBN) configurations with minimal parameterization, including the number of modules $B\in\left[1,\infty\right)$, the inter-module noise $\mu\in\left[0,1\right]$, the level of nested order within modules $p\in\left[0,1\right]$, and a shape parameter controlling the slimness of the nested structure $\xi\in\left[1,\infty\right]$. Illustratively, the nested matrix depicted in Fig.~\ref{fig:IBNMat} (a) corresponds to $B=1$, $p=0.3$, and $\mbox{connectance}=0.1$ ($\xi = 2.5$); the modular (b) and IBN matrices (c) correspond to $B=4$, $\mu=0.15$, and $\mbox{connectance}=0.1$, with the distinction that $p=1$ in the modular case while $p=0.1$ in the IBN case.

In addition to their internal structure, the weights of both competitive and mutualistic interactions are expressed as the result of a small perturbation, $\sigma^{\gamma}_{ij}$, around a mean value:
\begin{equation}
\Gamma^{AP}_{ij}=g_{ij}a_{ij},\quad g_{ij}=\gamma+\sigma^{\gamma}_{ij}\geq0,\quad \sigma^{\gamma}_{ij}\ll\gamma,
\end{equation}
where $a_{ij}=1$ if there exists a link between species $i$ and $j$, and zero otherwise. Matrices $\Gamma^{PA}$, $\Omega^{AA}$, and $\Omega^{PP}$ are treated similarly, with the mean competition value denoted as $\omega$. This approach aligns with previous literature \cite{allesina2012stability, May1973, Saavedra2015, Suweis2013} and allows for a uniform treatment of interaction strengths, assuming they are of the same order of magnitude for all species pairs, for both mutualistic and competitive weights. Consistent with the aforementioned studies, the interaction weight between the same species is set to one ($M_{ii}=1$).

\subsection{Stability}

Over the past few decades, a wide range of metrics designed to assess stability has emerged \cite{Grimm1997BabelOT, kefi2019advancing}. Here, we focus on the classical concept of stability, defined as the system's capability to restore the original equilibrium state after an infinitesimal perturbation of abundances.

This is evaluated by looking into the Jacobian matrix of the generalized Lotka-Volterra model,
\begin{equation}\label{JacDef}
J_{ij}\equiv\left(\frac{\partial \dot{x}_{i}}{\partial x_{j}}\right)_{\textbf{x}=\textbf{x}^*},
\end{equation}
where $\textbf{x}^*$ represents the stationary state of Eq.~\eqref{GeneralizedLotkaVolterraEq}, defined by the condition $ \dot{x}^*_i=0 $ and leading to 
\begin{equation}\label{StatSol}
\textbf{x}^*=M^{-1}\textbf{r},
\end{equation}
which describes the species abundances at equilibrium. 
Replacing Eq.\ \eqref{GeneralizedLotkaVolterraEq} into Eq.\ \eqref{JacDef} and recalling Eq.\ \eqref{StatSol} we reach the final form of the Jacobian matrix of the Lotka-Volterra model:
\begin{equation}\label{JacExp}
J_{ij}=-x^*_{i}M_{ij}.
\end{equation}
The system is stable if the real part of the largest eigenvalue of the Jacobian matrix is negative, otherwise is said to be unstable. Hence, the quantity 
\begin{equation}\label{lambda_star}
\lambda^*= -\text{Max}\left[Re\left(\lambda_J\right)\right],\quad \lambda_J\in\text{Sp}(J)
\end{equation}
naturally describes the stability and can be used to assess how different systems compare regarding such property.

In general, the stationary abundances, $x^*_i$, affect the expression of the Jacobian and have to be taken into account to obtain the eigenvalues, as seen in Eq.\ \eqref{JacExp}. Following the approach presented in \cite{Gibbs2018}, we assume that the abundances are all positive, \textit{i.e.} $x^*_i>0$ $\forall i$ which describes a context equivalent to sampling a proper $\textbf{r}$ vector in the feasibility domain.

\subsection{Feasibility}\label{sec:Feasibility}
In broad terms, feasibility refers to the system's capability to prevent extinctions and so maintain diversity, despite external perturbations, in the long-term limit. Formally, this translates into strictly positive stationary populations, \textit{i.e.} $x^*_i>0$ $\forall i$. In the case of the generalized Lotka-Volterra dynamics~\eqref{GeneralizedLotkaVolterraEq}, given a specific interaction matrix $M$, the emergence of stationary positive abundance depends solely on the self-growth rates $\textbf{r}$ \cite{Saavedra2015}. Herein, the goal is to characterize the range of possible growth rates associated with positive stationary abundances once the interaction matrix is provided. This is precisely what feasibility quantifies.

This problem is non-trivial from a mathematical standpoint and has attracted significant attention in recent years \cite{Svirezhev1983,rohr2014structural, Saavedra2015,grilli2017feasibility}. Particularly, it has been shown \cite{Ribando2006MeasuringSA, grilli2017feasibility} that the cumulative function of a multivariate normal distribution with mean value equal to zero and variance matrix $\Sigma^{-1}=2M^{t}M$:
\begin{equation}\label{OmegaDef}
\Theta = \frac{1}{(2\pi)^{S/2}\sqrt{\text{det}(\Sigma)}}\int ... \int_{\mathbb{R}^{S}\geq 0}e^{-\frac{1}{2}\textbf{x}^{t}\Sigma^{-1}\textbf{x}}d\textbf{x},
\end{equation}
constitutes a measure of the amount of growth rates associated with positive stationary abundances. The quantity $0\leq\Theta\leq1$ may be interpreted as the probability of randomly sampling an $\textbf{r}$ vector driving to positive abundances. Usually, scholars look into $\mathcal{F}\equiv\log_{10}\left(\Theta\right)$ which $-\infty<\mathcal{F}\leq0$, that is what is commonly referred as feasibility.

The higher the value of $\mathcal{F}$, the broader the range of growth rates associated with positive stationary abundances. In simpler terms, a higher $\mathcal{F}$ indicates a lower likelihood of extinctions in response to changes in growth rates. In this context, studying feasibility complements stability analysis by delving deeply into ecosystem persistence. While the latter concentrates on perturbations in abundances, the former considers changes in growth rates, encompassing all possible variables of Lotka-Volterra dynamics for a specific interaction matrix.


\section{Supplementary Material}
Supplementary material is provided in a separate file.

\section{Funding}
AL, AS-R and JB-H acknowledge financial support from the Spanish Ministerio de Ciencia e Innovaci\'on, through project No. PID2021-128966NB-I00. J.B-H. acknowledges financial support from the Ram\'on y Cajal program through the grant RYC2020-030609-I. AL also acknowledges financial support from the Spanish Ministerio de Ciencia e Innovaci\'on, through project No. PID2022-141802NB-I00 (BASIC).

\section{Author contributions statement}
AL, MP, AS-R and JB-H contributed equality to the work.

\section{Data availability}
The data underlying this article is publicly available in \cite{schwarz2020temporal}.


\end{document}